\documentclass[aps,preprint,preprintnumbers,amsmath,amssymb]{revtex4}
\usepackage{mathrsfs}
\usepackage{amssymb}
\usepackage{graphicx}
\usepackage{dcolumn}
\usepackage{bm}
\usepackage{xcolor}
\usepackage[breaklinks=true,colorlinks=true,linkcolor=blue,urlcolor=blue,citecolor=blue]{hyperref}
\usepackage{soul}
\usepackage{ulem}
\makeatletter

\newcommand{\Rmnum}[1]{\expandafter\@slowromancap\romannumeral #1@}

\begin{document}

\title{Skyrmion lattice creep at ultra-low current densities}

\author{Yongkang Luo$^{1,2}$}
\email[]{mpzslyk@gmail.com}
\author{Shi-Zeng Lin$^{1}$}
\author{Maxime Leroux$^{1}$}
\author{Nicholas Wakeham$^{1}$}
\author{David M. Fobes$^{1}$}
\author{Eric D. Bauer$^{1}$}
\author{Jonathan B. Betts$^{1}$}
\author{Joe D. Thompson$^{1}$}
\author{Albert Migliori$^{1}$}
\author{Marc Janoschek$^{1,3}$}
\email[]{marc.janoschek@psi.ch}
\author{Boris Maiorov$^{1}$}
\email[]{maiorov@lanl.gov}
\affiliation{$^1$ Los Alamos National Laboratory, Los Alamos, New Mexico 87545, USA;}
\affiliation{$^2$ Wuhan National High Magnetic Field Center and School of Physics, Huazhong University of Science and Technology, Wuhan 430074, China;}
\affiliation{$^3$ Laboratory for Neutron and Muon Instrumentation, Paul Scherrer Institute, 5232 Villigen PSI, Switzerland.}

\date{\today}

\maketitle

\textbf{Magnetic skyrmions are well-suited for encoding information because they are nano-sized, topologically stable, and only require ultra-low critical current densities $j_c$ to depin from the underlying atomic lattice. Above $j_c$ skyrmions exhibit well-controlled motion, making them prime candidates for race-track memories. In thin films thermally-activated creep motion of isolated skyrmions was observed below $j_c$ as predicted by theory. Uncontrolled skyrmion motion is detrimental for race-track memories and is not fully understood. Notably, the creep of skyrmion lattices in bulk materials remains to be explored. Here we show using resonant ultrasound spectroscopy--a probe highly sensitive to the coupling between skyrmion and atomic lattices--that in the prototypical skyrmion lattice material MnSi depinning occurs at $j_c^*$ that is only 4 percent of $j_c$. Our experiments are in excellent agreement with Anderson-Kim theory for creep and allow us to reveal a new dynamic regime at ultra-low current densities characterized by thermally-activated skyrmion-lattice-creep with important consequences for applications.
}\\

\textbf{Introduction}

In materials that exhibit dynamics under the application of external forces, the onset of motion is determined by the underlying pinning landscape. At zero temperature, a well-defined depinning threshold $j_c$ exists below which no motion arises, whereas far above $j_c$ linear dynamics occur [blue curve in Fig.~1(a)]. In contrast, at finite temperatures, motion may arise at a much lower threshold $j_c^*$ [red line in Fig.~1(a)] due to thermally-activated creep[Fig.~1(b)]. Creep is technologically-relevant in systems ranging from structural materials exposed to long-term mechanical stress\cite{Courtney-MaterMechnical}, to wetting front motion on heterogeneous surfaces\cite{Alava-2007}, to dynamics of domain walls\cite{Cayssol-DomainCreep}, and to vortex-motion in superconductors\cite{Larkin-JLTP1979,blatterrev}.

Early on theory suggested that creep may also be important for magnetic skyrmions\cite{Lin-particle}. Skyrmions are topologically-stabilized objects with a whirl-like spin-texture that emerge from competing magnetic interactions\cite{Muhlbauer-SKX}. Due to their topological stability and solitary particle-like behavior, they pin weakly to defects in the underlying atomic lattice, and substantially lower current densities $j$ are required to move skyrmions compared to magnetic domains. In turn, skyrmions are promising for applications in spintronics and racetrack memory devices based on controlled motion of particle-like magnetic nanostructures\cite{Fert,Mueller-2track}. Uncontrolled creep motion would be detrimental for devices making the understanding of skyrmion pinning crucial.

Indeed, creep motion of solitary skyrmions in thin films has been already observed\cite{JiangW-HallCreep}. However, in bulk materials, where skyrmions typically form a hexagonal skyrmion lattice (SKX) oriented in a plane perpendicular to an external magnetic field (\textbf{H}) [see Fig.~2(a)], creep remains elusive. This raises the question whether the pinning mechanisms in thin films and bulk materials are fundamentally different or if this merely due to differences in pinning landscape. As justified by experiments on both thin films and bulk materials \cite{JiangW-HallCreep, Schulz-MnSiTHE,Jonietz-MnSiCurrent}, skyrmion motion is described by weak collective pinning models\cite{Lin-particle, Reichhardt-SKXCreep}. For bulk materials, the critical current density ($j_c$), which denotes the onset of movement is expected to be determined by a trade-off between the strength of the pinning potential [$V(x)$, cf Fig.~1(b)] and the SKX stiffness\cite{blatterrev}. Distortion of the SKX in response to pinning centers with a characteristic \textit{Larkin length} $\xi$ [Fig.~1(c)] facilitates pinning. Because a perfectly rigid SKX cannot be pinned, $j_c$ depends {\it inversely} on the stiffness.

The application of a driving current tilts the pinning potential. At zero temperature, skyrmions escape only when $\Delta U$ $\leq$0 [Fig.~1(b)], where $\Delta U$ is the height of the tilted $V(x)$. In contrast, for finite temperature, the escape rate from pinning sites due to thermal fluctuations is given by $\Gamma\propto\exp{(-\Delta U/k_BT)}$, where $k_B$ is Boltzmann constant. When the external drive is small, a thermally-activated skyrmion follows an orbital trajectory around the pinning site due to the Magnus force [see Fig.~1(d)]. For $j_c^*$$<$$j$$<$$j_c$, a skyrmion may escape from a pinning center due to thermal fluctuations; however, it will immediately be trapped by a nearby pinning center\cite{Lin-particle}, resulting in creep [Fig.~1(e)]. Here $j_c^*$ is a threshold current that denotes that the energy barrier $\Delta U$ of the \textit{local} pinning potential is smaller than the energy $k_B T$ of thermal fluctuations. Effectively, skyrmions hop between pinning centers aided by thermal fluctuations, resulting in non-linear motion\cite{Reichhardt-SKXCreep}. The exact creep path depends on the type and distribution of such pinning centers. For SKX thermal fluctuations may depin small fractions of the lattice [shown in light yellow in Fig.~1(e)], where each fraction encounters a distinct local pinning landscape, in turn resulting of incoherent movement characteristic of \textit{local} creep. With increasing $j$, and thus decreasing $\Delta U$, the fraction of the SKX that may be depinned by fluctuation becomes larger. Eventually \textit{global} creep occurs, for which the entire SKX becomes depinned and repinned continuously due to a small but but non-zero $\Delta U$ that allows to recapture the SKX. For $j$$\gg$$j_c$, which denotes that  $\Delta U < 0$, the SKX flows freely through pinning centers [Fig.~1(f)], leading to a linear regime with coherent motion of the entire SKX [Fig.~1(a)].

The observation of creep motion of solitary skyrmions in thin films was achieved via polar magneto-optical Kerr effect (MOKE) microscopy using current pulses\cite{JiangW-HallCreep}. MOKE also was used to probe skyrmion diffusion, which is interesting in its own right as it has potential for applications in probabilistic computing, was recently observed in a thin film heterostructure that was engineered to have low  $\Delta U$ \cite{JZazvorka-2019}. Thermal diffusion of skyrmions can arise at \textit{zero current} at elevated temperatures that entail a thermal energy $k_B T$ larger than the unaltered energy barrier $\Delta U$($j$=0). However, the spatial resolution of MOKE is not sufficient to probe SKX motion, because skyrmions in bulk materials are substantially smaller (~10-100~nm) than in the thin films (~$\mu$m).

Instead, in bulk materials, microscopic skyrmion movement under current has been investigated by Lorentz Transmission Electron Microscopy (LTEM)\cite{Yu-FeGe} and via spin-transfer torque over the lattice by small-angle neutron scattering (SANS)\cite{Jonietz-MnSiCurrent}. Macroscopically, it has also been inferred from a reduction of the topological Hall effect (THE)\cite{PinningSKX,dong2015}. However, apart from the associated macroscopic $j_c$, no detailed information on the depinning process is available. In particular, the Hall effect is not suited to identify creep because it provides skyrmions with sufficient time to relax into equilibrium positions on the pinning sites, and thus eliminates side-jump motion caused by the Magnus force\cite{Reichhardt-SKXCreep}.

 To reveal previously elusive SKX creep in bulk materials we exploit the extreme sensitivity of Resonant Ultrasound Spectroscopy (RUS) to the coupling between the SKX and the atomic lattice\cite{LuoY-MnSi2018}. RUS probes the resonant frequencies $F_i$ of a solid which depend on its elasticity to determine the complete elastic tensor ($\mathbf{C}$). Details of how $\mathbf{C}$ is computed from the measured $F_i$ are described in \textbf{Supplementary Information (SI)}. RUS experiments under applied current on the prototypical SKX material MnSi allow us to probe the depinning of SKX with unprecedented resolution and thus to determine $j_c^*$ directly [see Fig.~2(a) and \textbf{Methods}]. We find that in MnSi creep motion occurs at a critical current density $j_c^*$ that is only 4 percent of $j_c$. We show that our experimental results are in excellent agreement with Anderson-Kim theory for creep\cite{Anderson-RMP1964} and connect the creep motion of skyrmion lattices in bulk materials with the previously known creep dynamics in thin films.

\textbf{Results}

\textit{Elastic Response to the Formation of the Skyrmion Lattice}

First, we briefly discuss RUS experiments at $j$=0, which establish the magnetic phase diagram [Fig.~2(b)], consistent with AC magnetic susceptibility ($\chi'$) (SI) and literature\cite{Muhlbauer-SKX}. The SKX phase appears in a narrow range of temperatures and field within the conically (CO) ordered magnetic phase. A fluctuation-disordered (FD) region\cite{Janoschek-MnSi2013} can also be seen right above $T_c$=28.7 K where the system undergoes a paramagnetic-helimagnetic (PM-HM) transition. In the absence of magnetic field, due to the cubic symmetry of MnSi, only three independent elastic moduli $C_{11}$, $C_{12}$ and $C_{44}$ are required. This is corroborated by our measurements [see Figure S4(a-c) in \textbf{SI}]. Under magnetic field $\mathbf{H}$ $\parallel$[001], the symmetry of the elastic tensor is lowered to tetragonal (\textbf{SI}), requiring three additional independent elastic moduli, $C_{33}$, $C_{23}$(=$C_{31}$) and $C_{66}$. In Figs.~3(a-c), we plot $C_{ij}$ as a function of $H$ at $T$=28 K. Each subset of $C_{ij}$ splits into two branches under magnetic field. We observe a discontinuous jump in some $C _{ij}$ between $H_{a1}$=1.4 kOe and $H_{a2}$=2.2 kOe. Based on the field dependence of $\chi'$ shown in Fig.~S2(b) (\textbf{SI}), we determine $H_{a1}$ and $H_{a2}$ as the lower and upper boundaries of the SKX phase, respectively. Below $H_{a1}$, there is a weak inflection in $C_{ij}(H)$ near $H_{c1}$=1.0 kOe, assigned as the field-induced HM-CO phase transition. Note that the elastic response to the SKX phase has not been seen previously\cite{Nii-MnSiElastic,Petrova-MnSiSKX} in off-diagonal moduli $C_{ij}$ ($i$$\neq$$j$). Because RUS probes all $C_{ij}$ in a single frequency sweep\cite{Migliori-RUS,Migliori-PhysicaB1993}, it directly reveals the shear modulus $C^*$$\equiv$$(C_{11}$$-$$C_{12})/2$,  which exhibits a much smaller variation than the compression moduli. Further, the jump in $C^*$ is an order of magnitude bigger than in $C_{66}$, indicating that hexagonal symmetry of the SKX does not describe the system, as this requires $C^*$=$C_{66}$ (\textbf{SI}). Instead, this shows that RUS probes the response of the chemical lattice (which is tetragonal in field) to the formation of the hexagonal SKX.

We note that when tracking changes and discontinuities, it is more reliable to plot the raw frequencies\cite{Evans2017}. As shown in Table S1 in the \textbf{SI}, the two resonances $F_{1654}$ and $F_{2419}$ are predominantly related to $C_{11}$.  For a frequency that depends only on one $C_{ij}$ we have $C_{ij}$$\propto$$F^2$, so for small changes in $C_{ij}$ we have $\delta C_{ij}$ $\sim$2$\delta F$. Fig.~3(e) displays the temperature dependence of $F_{2419}$ at various magnetic fields. For $H$=0, the profile of $F_{2419}(T)$ resembles that of $C_{11}(T)$ [Fig.~S4(a)], confirming the dominance of $C_{11}$. With increasing magnetic field, $T_c$ is gradually suppressed, and the signature of the phase transition becomes more pronounced for $H$$>$$2.5$ kOe. $T^*$, the temperature where the minimum of $F_{2419}(T)$ occurs, initially decreases with increasing $H$ but then broadens and shifts to higher $T$ for $H$$>$$2.5$ kOe where spins become polarized by the external field. The window between $T^*$ and $T_c$ describes the FD region in Fig.~2(b). Figure~3(f) shows $F_{1654}$ as a function of $H$ measured at selected temperatures. The discontinuous jump in $F_{1654}(H)$ can be identified between 26.9 K and 28.5 K similarly as seen in $C_{11}$ in Fig.~3(a). A positive jump in $F_{1654}(H)$ signifies stiffening in $C_{11}$ when the system enters the SKX phase. The (maximal) magnitude of the jump in $F_{1654}(H)$, denoted by $\Delta F$ [see Fig.~4(e)], is plotted as a function of temperature in Fig.~4(g) with a maximum near the SKX-FD boundary and decreasing as $T$ decreases. The value of $\Delta F$, therefore, is a qualitative measure for the coupling between chemical lattice and SKX. Indeed, the maximum in $\Delta F$ near the upper boundary of the SKX ($T$$\sim$28.2 K) phase can be explained by proximity to the FD regime (see below).

\textit{Elastic Response under Applied Current}

Now that we have established the elastic response to the presence of the SKX, we demonstrate the influence of applied electrical current. Figures~4(a-d) display the field dependence of $F_{1654}$ for various applied electrical currents (\textbf{I}$\perp$\textbf{H}) at four selected temperatures 28.4, 28.2, 27.7 and 27.1 K inside the SKX phase, respectively. We note that the resonance $F_{1654}$ is most sensitive to the presence of the SKX because it is related to the elastic moduli $C_{11}$ which is more than an order of magnitude larger than any other $C_{ij}$. In addition, $F_{1654}$ provides the highest quality signal as described in more detail in the \textbf{SI}. Taking $T$=27.7 K as an example [Fig.~4(c)], $\Delta F$ remains essentially unchanged for $j$ up to 56.5 kA/m$^2$, drops abruptly between 59.1 and 64.5 kA/m$^2$, and becomes nearly unresolvable at 67.2 kA/m$^2$, as if the SKX is completely decoupled from the lattice. The threshold current density $j_c^*$ is defined as the midpoint of the drop in $\Delta F$, shown in Fig.~4(f) with $j_c^*$=62(3) kA/m$^2$ at 27.7 K. The error bar is set by the step size in current. In Fig.~4(f) we also show $\Delta F(j)$ for 27.1 K, and the difference in $j_c^*$ is far larger than the measurement uncertainty. We emphasize that the changes observed in the resonance frequencies are not caused by current-induced Joule heating as can be illustrated by several observations. (i) The magnitude of $\Delta F$ does not increase with $j$ [cf Fig.~4(f)] for temperatures near the lower boundary of the SKX phase (e.g. 27.1 K); (ii) $j_c^*$ does not increase monotonically with decreasing temperature; (iii) the phase boundaries in $H$ depend strongly on temperature as revealed in Fig.~2(b), but the width of the SKX phase with respect to field showing non-zero $\Delta F$ remains unaffected for increasing $j$ at all measured temperatures, as expected for constant temperature [Figs.~4(a-e)].

\textbf{Discussion}

The drastic changes in the elastic properties above $j_c^*$ suggest that skyrmion depinning occurs at critical current densities that are a factor of 25 smaller than $j_c\sim$1.5 MA/m$^2$ derived from previous measurements\cite{Schulz-MnSiTHE,Jonietz-MnSiCurrent,Yu-FeGe,dong2015,Leroux-FeGeTHE}. There are several scenarios that may explain this disparity. First, the presence of ultrasonic waves could facilitate depinning by shaking skyrmions off pinning potentials yielding a smaller current for motion threshold as previously observed for superconducting vortices\cite{Valenzuela2001}. However, this effect is unlikely here because the same $j_c^*$ is observed with ultrasonic excitation with twice the amplitude [full symbols in Fig.~4(h)]. Another possibility is the difference in pinning defects in our sample compared to previous studies. However, a detailed characterization of our sample (see \textbf{SI}) reveals that it is of the same high-quality as samples used in previous studies\cite{Muhlbauer-SKX, Janoschek-MnSi2013, Schulz-MnSiTHE,Jonietz-MnSiCurrent}, as notably demonstrated by a large residual resistivity ratio ($RRR$=87).

As we discuss in the following, a consistent view on the difference between $j_c$ and $j_c^*$ may be established by considering the different sensitivity in detecting the onset of skyrmion motion with distinct techniques. As explained in the introduction, THE is unable to measure incoherent skyrmion motion resulting from creep. Similarly, SANS is only sensitive to coherent rotation of the entire SKX due to spin-transfer torque\cite{Jonietz-MnSiCurrent}. In contrast, RUS directly measures the magneto-crystalline coupling, and thus detects skyrmion movement immediately when the SKX decouples from the atomic lattice. Thus, it can detect motion due to creep at much lower current density [cf. Fig.~1(a)] as corroborated by the abrupt change in $\Delta F$ as $j$ reaches $j_c^*$ that is contrasted by the gradual decrease of the topological Hall resistivity for $j$$>$$j_c$ \cite{Schulz-MnSiTHE}. This is similar to superconducting vortices, where  magnetization measurements and electrical transport are sensitive to creep and flux-flow changes, respectively\cite{blatterrev,borisnatmat,campbellevettsnew}.

That $j_c^*$ marks the presence of a creep regime is also evidenced by the fits of our data to Anderson-Kim theory for creep for which the local pinning potential vanishes linearly with current $\Delta U$=$\beta(j$$-$$j_c^*)$, where $\beta$ is a pre-factor \cite{Anderson-RMP1964}. Notably, the measured $\Delta F$ is well-described by $\Delta F(j)$$-$$\Delta F$($j$=0) $\propto \exp[(j-j_c^*)\beta/k_B T]$ over the entire temperature range of the SKX phase [Fig.~4(f)]. This creep scenario is further in agreement with the observed temperature dependence of $j_c^*$. As described above, according to weak collective pinning theory, $j_c^*$ is inversely proportional to the SKX stiffness. The temperature dependence of $\Delta F$($j$=0) [see Fig.~4(g)] displays two trends. Starting at high temperature from the SKX-FD boundary, $\Delta F$ initially increases as $T$ decreases becoming stiffer down to $T$=28.2 K where $j_c^*$ also minimizes [vertical arrows in Fig.~4(g) and (h)]. The behavior for $T$ $>$28.2 K is consistent with strong thermal fluctuations near the upper boundary of the SKX phase that soften the SKX lattice which allows to better accommodate local pinning sites, in turn, improving pinning. The resulting enhancement of $j_c^*$ near the SKX-FD phase boundary is called \textit{peak effect} and was also observed in MnSi via the THE\cite{Schulz-MnSiTHE} [see orange diamond symbols and line in Fig.~4 (h)] and is well-documented for superconducting vortices\cite{blatterrev,Valenzuela2002}. As $T$ continues to decrease, $j_c^*$ and $\Delta F$($j$=0) keep displaying an inverse relation consistent with weak collective pinning down to $T$=27.7 K where $j_c^*(T)$ shows a maximum but $\Delta F(T)$ is featureless.

To understand the behavior below $T$=27.7 K, it is important to consider that for bulk materials such as MnSi, the size and shape of the SKX phase is sensitive to the sample geometry due to demagnetization effects \cite{Bauer_MnSi_sampleshape}. In addition, it has been shown via neutron diffractive imaging that the phase transition to the conical phase is characterized by macroscopic phase separation where only parts of the sample show SKX order, whereas the rest exhibits conical order \cite{Reimann-MnSinDI}. In addition,where in the sample the SKX phase nucleates (edge vs. center) strongly varies as a function of magnetic field \cite{Reimann-MnSinDI}. For plate-like samples as were required for our combined RUS and current study, the influence of demagnetization fields is particularly strong in the part of SKX phase that is characterized by $T<$ 27.7 K as the macroscopic phase separation is observed for more than 50\% of the field range of the SKX phase \cite{Bauer_MnSi_sampleshape, Reimann-MnSinDI}. The fraction of the SKX phase showing phase separation increases further when the temperature is lowered\cite{Bauer_MnSi_sampleshape, Reimann-MnSinDI}. Naturally, the prominent phase separation in the low-temperature regime of the SKX phase, results in magnetic domain boundaries between the conical and SKX phases, which influences the pinning of the SKX. This distinct pinning regime is reflected in a change of the behavior of $j_c^*(T)$ for $T<$ 27.7 K.

It is interesting to compare the SKX creep in the bulk material MnSi identified here to the creep of individual observed in thin films reported previously \cite{JiangW-HallCreep}. The ratio of $j_c/j_c^*$ found for MnSi varies between 10 and 50 depending on temperature, which is a factor 10-20 larger than the ratio in thin films. Because both the onset of creep at $j_c^*$ and the onset of coherent motion at $j_c$ depend on the pinning potential, it is unlikely that this difference is due to a difference in amount and nature of the defects that pin skyrmions in bulk and thin film materials, respectively. Instead this difference originates from a skyrmion lattice being easier to recapture by pinning sites compared to single skyrmions, and therefore a substantially larger $j_c/j_c^*$ is needed to enter the linear driven regime. Finally, in contrast to single skyrmions, for SKX we can also differentiate local and global creep motion. As discussed above $j_c^*$ denotes the onset of local creep where only part of the SKX is depinned. Our RUS measurements demonstrate that for current densities $j$ that are a only few percent larger than $j_c^*$, $\Delta F$ vanishes. Because $\Delta F$ is a measure of coupling between the SKX and the underlying lattice, this suggest a crossover from local to global creep, where on average the SKX is depinned, but is recaptured continuously.

In conclusion, our RUS measurements on the prototypical SKX material MnSi under applied electrical currents provide evidence for the existence of a novel regime of skyrmion lattice dynamics in bulk materials at substantially lower depinning current densities $j_c^*$ than previously reported for thin films. The temperature dependence of this new intermediate regime is consistent with thermally induced creep of a skyrmion lattice. Our results directly connect the creep motion of skyrmion lattices in bulk materials with the known creep dynamics in thin films, showing that the underlying assumptions for a tilted local weak pinning potential\cite{Lin-particle, Reichhardt-SKXCreep} is the correct model for both cases despite obvious differences in the interactions that support the emergence of skyrmions. This highlights that our current theoretical understanding of skyrmion creep is complete and will be  relevant for applications. Notably, because $j_c^*$ is only about four percent of $j_c$ above which coherent skyrmion motion occurs, it is crucial that any devices based on the control of skyrmion dynamics must be carefully engineered to avoid uncontrolled creep. This is particularly critical as real devices will have to work at room temperature, where thermally activated creep will be substantial.

\textbf{Methods}

The MnSi single crystal was grown by the Bridgman-Stockbarger method followed by a 1-week anneal at 900$^\circ$C in vacuum. The stoichiometry of the crystal was examined by energy dispersive x-ray spectroscopy (EDS). The sample was polished into a parallelepiped along the [001] direction with dimensions 1.446$\times$0.485$\times$0.767 mm$^3$. The orientation of the crystal was verified by Laue X-ray diffraction within 1$^\circ$. Electrical resistivity and AC susceptibility measurements revealed a magnetic transition as expected at $T_c$=28.7 K, and a residual resistance ratio $RRR$[$\equiv\rho(300~\text{K})/\rho(T$ $\rightarrow$0)]=87, indicating a high quality single crystal [See Fig.~S1 in \textbf{SI}]. All the measurements in this work were performed on the same crystal. Further, the SKX in a different piece of this sample was directly observed using SANS\cite{Fobes-MnSiStrain}.

A schematic diagram of the RUS setup is shown in Fig.~2(a). The sample was mounted between two LiNbO$_3$ transducers. In order to stabilize the sample in a magnetic field and maintain RUS-required weak transducer contact, Al$_2$O$_3$ hemispheres (that also act as wear plates and electrical insulators) were bonded to each transducer. The external magnetic field $\textbf{H}$ was applied along [001] of the cubic crystal structure of MnSi. Frequency sweeps from 1250 to 5300 kHz were performed for each measurement. The resonance peaks were tracked and recorded as a function of temperature and magnetic field. Elastic moduli $C_{ij}$ were extracted from 24 resonance frequencies with an RMS error of 0.2\% using an inversion algorithm\cite{Migliori-RUS,Migliori-PhysicaB1993}. Although the absolute error is large (mainly because of uncertainties in the sample dimensions), the precision of the elastic moduli determination is at least 1$\times$10$^{-7}$.  Finally, we note that by measuring all the elastic moduli $C_{ij}$ {\it simultaneously} with a fixed magnetic field orientation, accounting for anisotropic demagnetization factors is unnecessary and direct comparisons among $C_{ij}$ can be made.

To study the effect of current on moduli, we attached gold wires (13 $\mu$m) at opposite sides of the specimen allowing a DC current to be applied along [100], \textbf{I}$\perp$\textbf{H} with a cross-section of 0.485$\times$0.767 mm$^2$. To minimize Joule heating at the contacts, the Au wires were spot-welded to the sample and covered with silver paint to improve current homogeneity and reduce contact resistance. The resulting Ohmic electrical contacts were less than 0.5 $\Omega$. Whenever the current was changed, we waited for steady state before recording. A small temperature increase ($<$30 mK) was observed at the thermometer right after applying relatively larger currents. We compensated for this by adjusting the set-point of the temperature controller.

\textit{Data Availability Statement (DAS)}: The data that support the plots within this paper and other findings of this study are available from the corresponding author upon reasonable request.

\textit{Code Availability Statement}: The algorithm used to fit the RUS data is detailed in the \textbf{SI}. The computer code itself is available from the corresponding author upon reasonable request.

\emph{}\\

\textbf{Acknowledgments}\\
We thank F.~F. Balakirev for technical support, and F. Ronning and M. Garst for insightful conversations. Sample synthesis by EDB and characterization by JDT was performed under the U.S. DOE, Office of Science, BES project "Quantum Fluctuations in Narrow Band Systems". Research by YL, SL, DMF, ML, ND, BM and MJ was supported by LANL Directed Research and Development program project "A New Approach to Mesoscale Functionality: Emergent Tunable
Superlattices (20150082DR)" [PI Janoschek]. Work by JB and AM was part of the Materials Science of Actinides, an Energy Frontier Research Center funded by the U.S. DOE, Office of Science, BES under Award DE-SC0001089. YL acknowledges support by the 1000 Youth Talents Plan of China.

\textbf{Author contributions}\\
YL, SL, MJ, and BM conceived and designed the experiments. EDB synthesized the samples. ML, NW, DMF and JDT characterized the crystals. JB and AM provided technical support for the experimental set-up. YL performed most of the ultrasonic measurements. YL, SL, MJ and BM discussed the data, interpreted the results, and wrote the paper with input from all the authors.

\textbf{Author information}\\
The authors declare no competing financial interests. Correspondence and requests for materials should be addressed to Y. Luo (mpzslyk@gmail.com), M. Janoschek (marc.janoschek@psi.ch) or B. Maiorov (maiorov@lanl.gov).

\newpage
\textbf{Figures:}

\begin{figure}[!h]
\hspace*{-10pt}
\includegraphics[width=16cm]{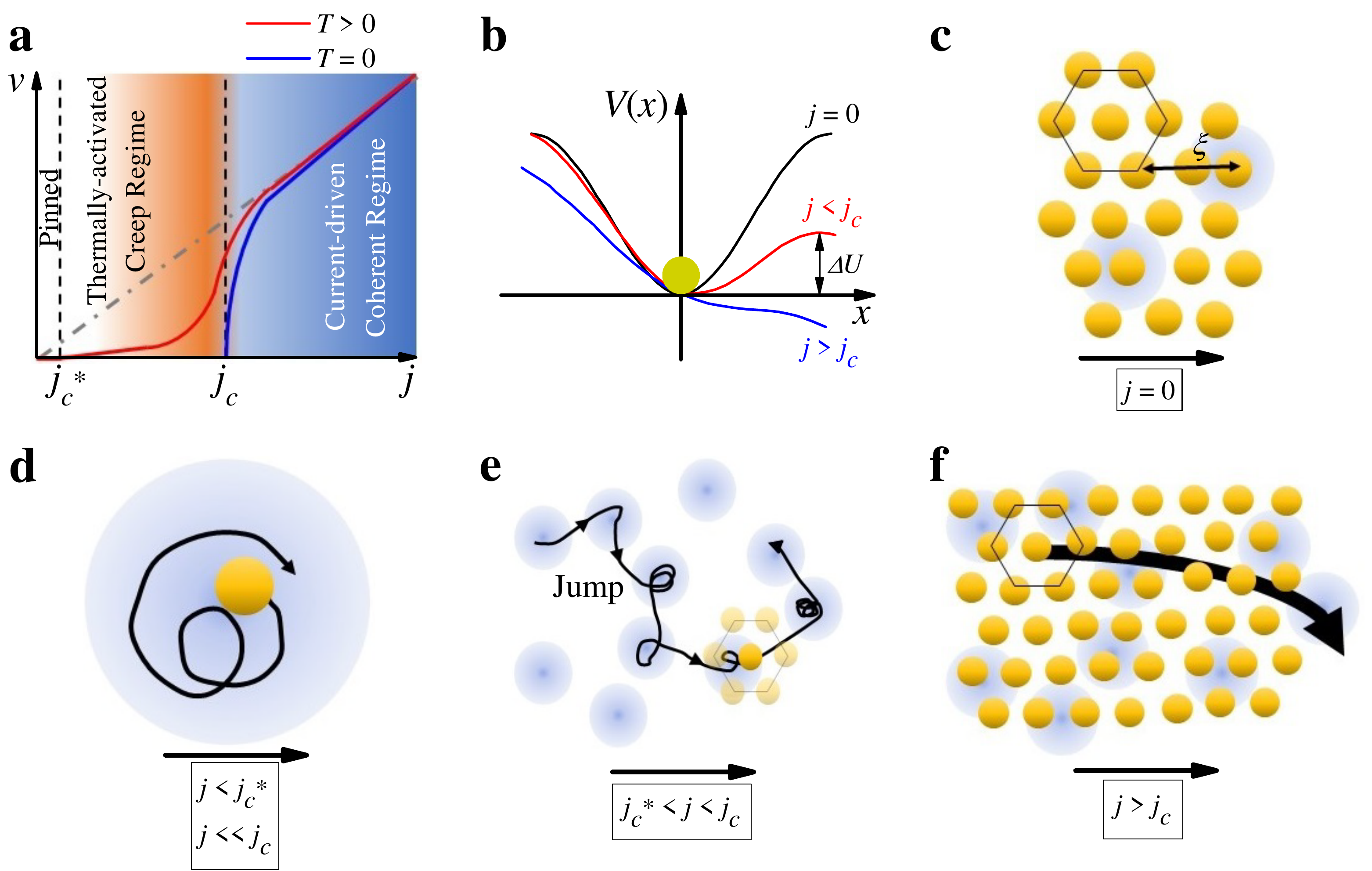}
\label{Fig1}
\end{figure}
\vspace*{0pt}
\textbf{Fig.~1 $|$ Skyrmion creep at ultra-low current densities.} (a) The dynamic response of skyrmions characterized by a competition of an external drive  $j$ (here $j$ is an applied current density) with a pinning potential $V(x)$ [see (b)] can be parametrized via its velocity $v$. This allows to identify three dynamic regimes that are illustrated in panels (d-f) and explained in the following.  (b) Local pinning potential $V(x)$ for different current densities. The applied current tilts the pinning potential. Above a critical current density $j_c$, the energy barrier $\Delta U$ vanishes completely allowing the skyrmion to depin. At zero temperature $T$ (blue line in (a)), this well-defined depinning threshold $j_c$ thus defines two dynamic regimes, where the skyrmion is either pinned or exhibits current-driven motion. In contrast, for finite $T$ [red curve in (a)], the tilted potential promotes thermal activation of skyrmions already for $j_c^*$$<$$j$$<$$j_c$, resulting in a third regime defined by creep. Here $j_c^*$ is the threshold when skyrmions start to depin from the \textit{local} pinning center, but remain globally pinned. (c) A skyrmion lattice (yellow dots) pinned to a few pinning centers (blue shaded areas) is shown. The skyrmion lattice is distorted to accommodate the pinning center, with a characteristic length $\xi$ known as the Larkin length. (d) For a small current density $j$$\ll$$j_c$ a thermally activated single skyrmion follows an orbital trajectory around the pinning site because of the Magnus force. (e) For larger currents $j$$<$$j_c$ creep may occur. Here the skyrmion spends most of the time orbiting the pinning potential. Further, when the skyrmion escapes from one pinning center due to thermal fluctuations, it will immediately be trapped by a nearby pinning center, resulting in creep motion. (f) For large current densities $j$$>$$j_c$, the skyrmion lattice flows freely through the pinning centers and the lattice order is improved as has been previously observed\cite{Yu-FeGe}. \\

\newpage

\begin{figure}[!h]
\hspace*{-10pt}
\includegraphics[width=14cm]{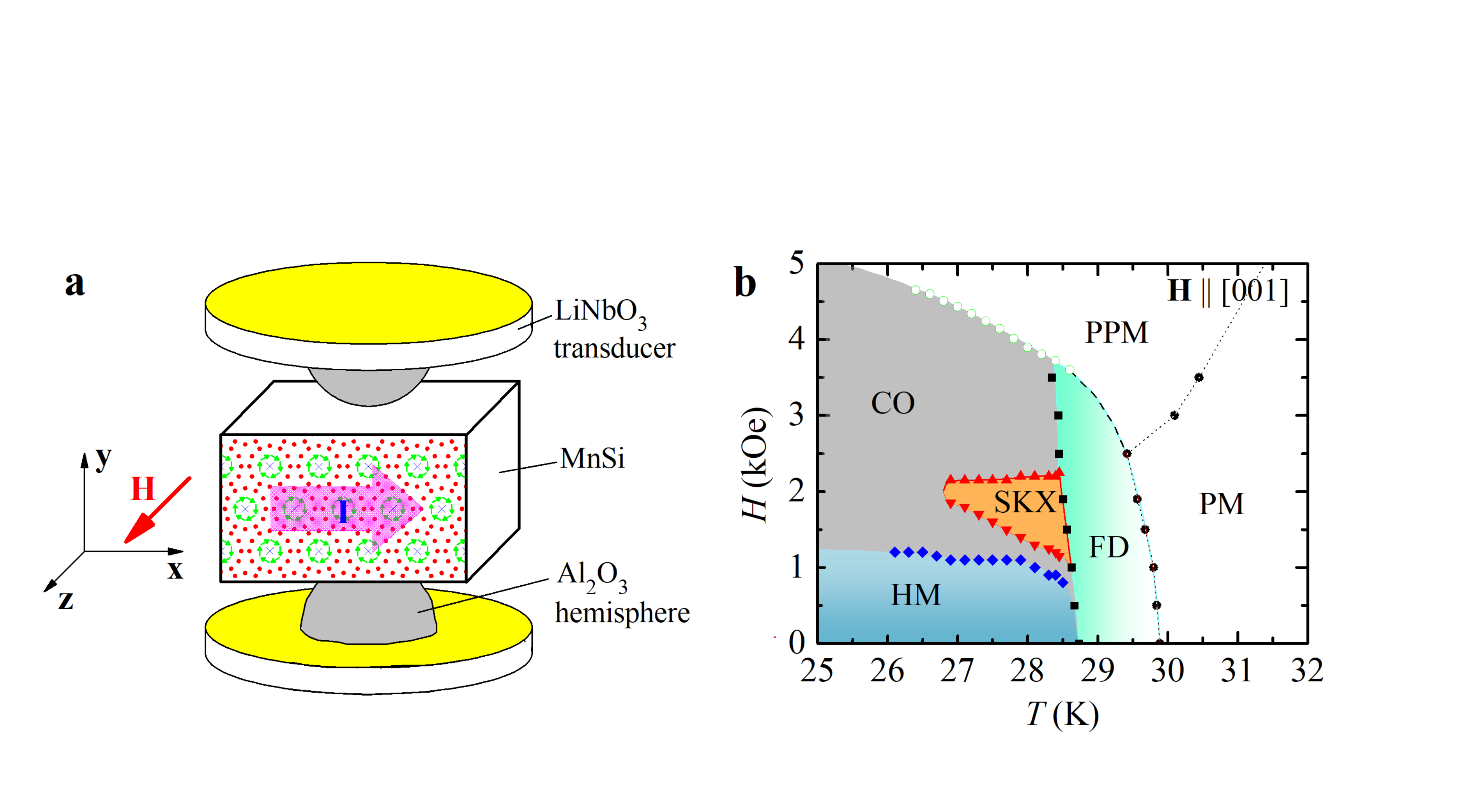}
\label{Fig2}
\end{figure}
\vspace*{-20pt}
\textbf{Fig.~2 $|$ Resonant ultrasound spectroscopy (RUS) experimental setup and phase diagram of MnSi.} (a) Schematic diagram of the RUS experimental setup. The sample is mounted between two LiNbO$_3$ transducers with the bottom one serving as ultrasonic driving source and the top one as pick-up. The magnetic field $\mathbf{H}$ is applied along [001], and the driving current $\mathbf{I}$ is along [100]. (b) Magnetic phase diagram established via RUS. The open circles are from $\chi'(H)$, defined as the mid-point of spin polarization. The abbreviations denoting the different phases are: HM = helimagnetic, CO = conical, SKX = skyrmion lattice, FD = fluctuation disordered, PM = paramagnetic, and PPM = polarized paramagnetic. \\

\newpage

\begin{figure}[!h]
\hspace*{-20pt}
\includegraphics[width=17cm]{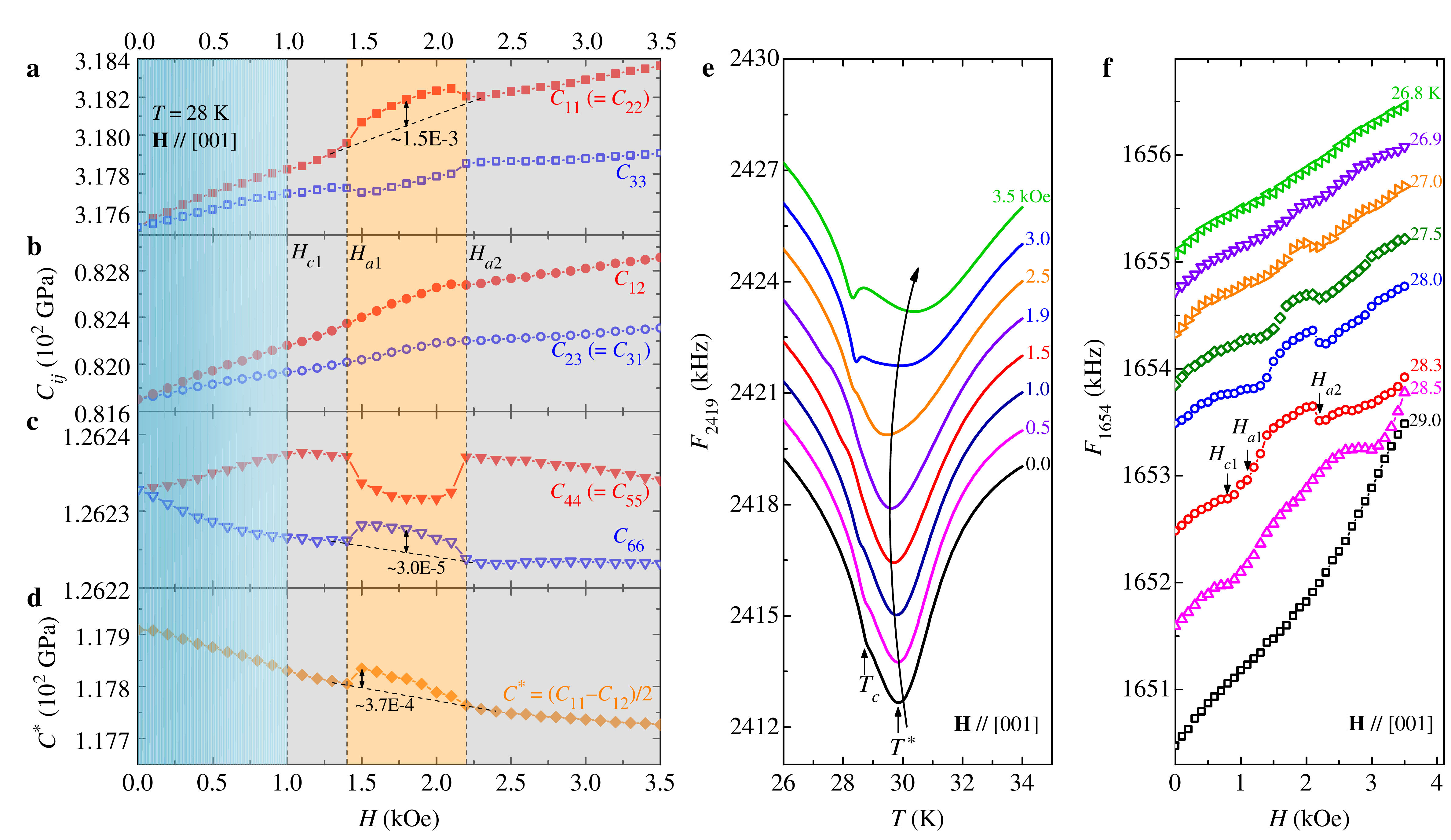}
\label{Fig3}
\end{figure}
\vspace*{-20pt}
\textbf{Fig.~3 $|$ Elastic properties of MnSi at $j$=0 determined by resonant ultrasound spectroscopy (RUS).} (a-d) Isothermal elements $C_{ij}$ and $C^*$$\equiv$($C_{11}$$-$$C_{12}$)/2 of the elastic tensor $\mathbf{C}$ as a function of magnetic field $H$, respectively. Although both compression ($C_{11}$, $C_{33}$) and shear moduli ($C_{44}$, $C_{66}$) display abrupt changes, $C_{12}$ and $C_{23}$ only exhibit a slight change in slope near $H_{a1}$ and $H_{a2}$. The accuracy of the absolute values of the $C_i$ is determined by the quality of the fits of the resonance frequencies, which is of the order of 0.1\% (see Ref.~\cite{Migliori-PhysicaB1993} and table S1 in \textbf{Supplementary Information (SI)}). However, our results rely on relative shifts of the $C_i$, which can be determined with an accuracy of 0.01\%.  (e) Temperature ($T$) dependence of the resonant frequency $F_{2419}$ measured for various $H$. $T_c$ is the critical temperature for the paramagnetic (PM) to helimagnetic (HM) transition, and $T^*$ is the characteristic temperature below which the fluctuation-disordered (FD) regime arises. (f) $F_{1654}$ as a function of $H$, measured at selected temperatures $T$. The accuracy of determining relative frequency shifts with our RUS setup is better than 0.01 \% \cite{Migliori-PhysicaB1993}.  \\

\newpage

\begin{figure}[!h]
\hspace*{-20pt}
\includegraphics[width=17cm]{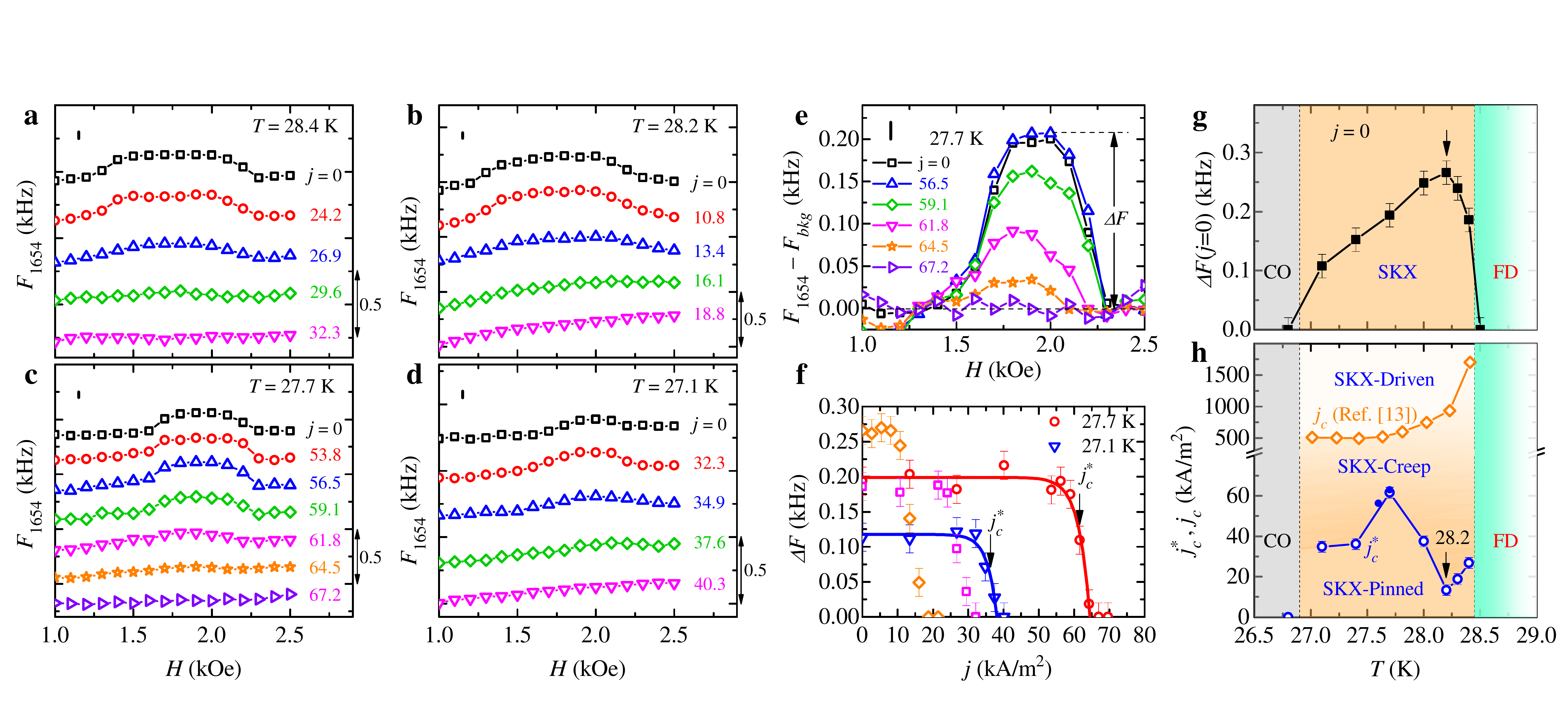}
\label{Fig4}
\end{figure}
\vspace*{-20pt}
\textbf{Fig.~4 $|$  Signatures of skyrmion creep in the elastic properties.} (a-d) Field dependence of the resonance frequency $F_{1654}$ at various current densities $j$, measured at temperatures $T=$28.4 K, 28.2 K, 27.7 K and 27.1 K, respectively. The curves are vertically offset for clarity. The double arrow on the right side of each panel denotes the scale of the frequency shift.  The accuracy of determining relative frequency shifts with our RUS setup is better than 0.01 \% \cite{Migliori-PhysicaB1993}. (e) $F_{1654}$$-$$F_{bkg}$ as a function of $H$ at 27.7 K, where $F_{bkg}$ is the smooth background of $F_{1654}$. $\Delta F$ is defined as the maximum of $F_{1654}$$-$$F_{bkg}$. (f) $\Delta F$ vs. $j$ for 27.1, 27.7, 28.2 and 28.4 K. The solid lines are fits to Anderson-Kim model, $\Delta F(j)$$-$$\Delta F(j$$=$$0)\propto \exp[\beta(j$$-$$j_c^*)/k_B T]$,  where $\beta$ is material dependent a pre-factor. In the Anderson-Kim model the local pinning potential vanishes linearly with current $\Delta U$=$\beta(j$$-$$j_c^*)$. The arrows marks the critical current density value $j_c^*$=62 kA/m$^2$ for 27.7 K to illustrate how critical current densities where determined. (g) $\Delta F$ as a function of $T$ in the absence of current. (h) Temperature dependent $j_c^*$ (blue line and symbols). The full symbols are measured with an ultrasonic excitation two times larger than that for the open symbols. The orange line and symbols are reproduced from Ref. \cite{Schulz-MnSiTHE} and denote the onset of coherent linear skyrmion lattice motion as determined by the reeduction of the topological Hall effect (THE) under current.\\

\newpage

\setcounter{table}{0}
\setcounter{figure}{0}
\setcounter{equation}{0}
\setcounter{section}{0}
\renewcommand{\thefigure}{S\arabic{figure}}
\renewcommand{\thetable}{S\arabic{table}}
\renewcommand{\theequation}{S\arabic{equation}}
\onecolumngrid

\begin{center}
{\it\textbf{Supplementary Information: }} \\
\textbf{Skyrmion creep at ultra-low current densities}\\

\end{center}

\begin{center}
Yongkang Luo$^{1,2*}$\email{mpzslyk@gmail.com}, Shizeng Lin$^{1}$, M. Leroux$^{1}$, N. Wakeham$^1$, D. M. Fobes$^{1}$,  E. D. Bauer$^{1}$, J. B. Betts$^{1}$, A. Migliori$^{1}$, J. D. Thompson$^1$, M. Janoschek$^{1,3\dag}$$\email{marc.janoschek@psi.ch}$, and Boris Maiorov$^{1\ddag}$\email{maiorov@lanl.gov}\\
$^1${\it Los Alamos National Laboratory, Los Alamos, New Mexico 87545, USA;}\\
$^2${\it Wuhan National High Magnetic Field Center and School of Physics, Huazhong University of Science and Technology, Wuhan 430074, China; and}\\
$^3${\it Laboratory for Neutron and Muon Instrumentation - Paul Scherrer Institut - Villigen PSI - Switzerland.}\\

\date{\today}
\end{center}

\vspace{10pt}

In this \textbf{Supplementary Information (SI)}, we provide additional electrical resistivity, magnetic AC susceptibility data and energy dispersive x-ray spectroscopy (EDS). We also give a brief introduction of elastic moduli and details of data analysis used for RUS measurements that support the results, discussions and our conclusions given in the main text. \\

\textbf{SI \Rmnum{1}: S\lowercase{ample characterization}}

Figure S1(a) shows the temperature dependence of resistivity $\rho$ measured with current applied along [100]. The resistivity decreases sub-linearly with decreasing $T$. Below 32 K, $\rho(T)$ starts to decrease rapidly. An inflection point is observed at 28.7 K [inset to Fig.~S2(a)], used to determine  $T_c$=28.7 K. We also use this data set to calculate the residual resistivity ratio ($RRR$) of our sample via $\rho(300~\text{K})/\rho(T$$\rightarrow$0)=87, which is typical for high quality MnSi crystals\cite{SMuhlbauer-SKX,SJanoschek-MnSi2013}. In the inset to Fig.~S1(a), we also display the derivative of the resistivity with respect to temperature, $d\rho/dT$, as a function of $T$ that typically tracks the specific heat. Indeed, a sharp peak in $d\rho/dT$ is found at $T_c$ on top of a broad maximum which extends to $\sim$32 K analog to previous specific heat measurements\cite{SBauer-MnSiC} as well as the thermal expansion\cite{SStishov-MnSi2007}. Here the broad maximum marks the onset of strong spin fluctuations due to a Brazovskii scenario that result in a fluctuation-induced first-order transition at $T_c$ \cite{SJanoschek-MnSi2013}.

In Fig.~S1(b) we present the results of AC susceptibility measured with an alternating field $H_{ac}$=3 Oe and frequency $f$=200 Hz. When $H_{dc}$=0, $\chi'(T)$ displays a pronounced peak at $T_c$, consistent with resistivity and RUS measurements. For $H$=1.9 kOe, $\chi'(T)$ shows a pronounced valley between 26.8 and 28.6 K characteristic of the SKX phase. This SKX phase is also seen in the isothermal $\chi'(H)$ curves shown in Fig.~S2(c). These observations are akin to earlier literature\cite{SBauer-MnSiPhase,SNii-MnSiElastic}, except for the slight different field range for SKX which is probably due to a different demagnetization factor. \\

\begin{figure*}[htbp]
\vspace*{-20pt}
\includegraphics[width=17cm]{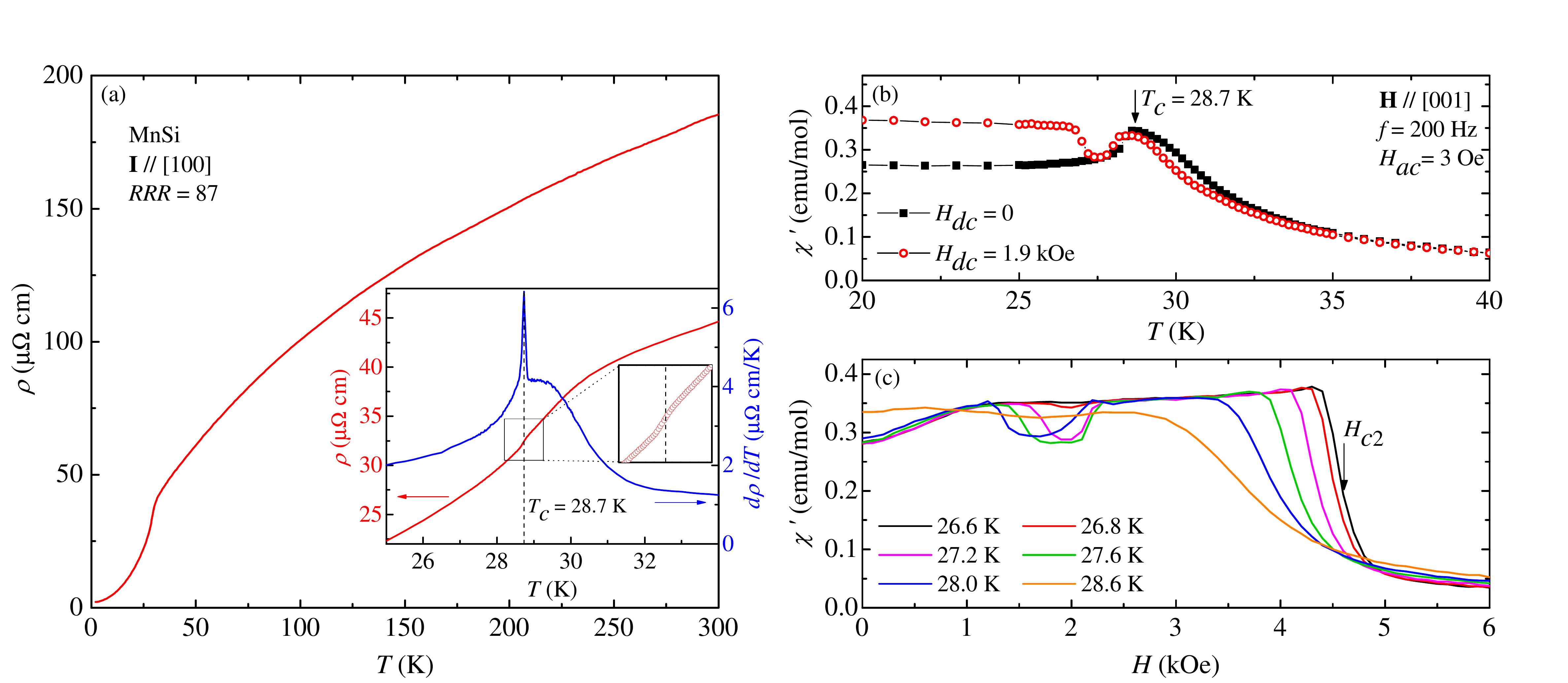}
\vspace*{-30pt}
\end{figure*}
\textbf{Fig.~S1 $|$ Resistivity and AC magnetic susceptibility of MnSi.} (a) Temperature dependence of resistivity. The inset shows an enlarged view of $\rho(T)$ as well as $d\rho/dT$ near $T_c$=28.7 K. (b) Temperature dependent AC susceptibility $\chi'$, measured under DC magnetic field $H$=0 (black) and 1.9 kOe (red). (c) $\chi'(H)$ measured at various temperatures. The SKX phase appears in the window 26.8 K$<$$T$$<$28.6 K and between 1.5 and 2.3 kOe. \\

Additionally, the stoichiometric concentrations of Mn and Si of the sample were determined by EDS measurements. Figure S2(a) shows a representative EDS spectrum. The distribution maps of Mn and Si are given in Fig.~S2(b) and (c), respectively. The average mole ratio Mn:Si=0.88:1.12. Note that the systematic uncertainty of EDS can be as large as 10\% \cite{SNewbury-EDS2013}, and in particular, the concentration of Mn is commonly underestimated\cite{SNewbury-EDS2015}. In light of these well-documented systematic errors of the EDS method, we conclude that the bulk of our crystal is stoichiometric, in particular because EDS is a surface sensitive probe. This is corroborated by all other measured properties that are consistent with reports in literature where even fine details are reproduced. Notably, we emphasize that even slight Mn deficiency of 2.2\% has been reported to decrease $T_c$ by 1.5~K \cite{STOu-Yang-2015}, which is not observed for our sample.  Furthermore, the same study reported that the shoulder in the specific heat signalling the onset of the fluctuation disordered (FD) regime (see Fig.~2 in main text) is washed out due to Mn deficiency\cite{STOu-Yang-2015}; in stark contrast, clear signatures of the FD regime are observed both in our RUS measurements, as well as in the derivative of the resistivity with respect to temperature, $d\rho/dT$, that typically tracks the specific heat. At the same time, 2.2 \% of Mn deficiency\cite{STOu-Yang-2015} was shown to reduce the residual resistivity ratio (RRR) from 85, which is typical for high quality MnSi crystals\cite{SMuhlbauer-SKX,SJanoschek-MnSi2013}, to 38. For our single crystal we observe $RRR$=87 (see Fig.~S1) suggesting that it is indeed stoichiometric. Finally, the shape of the SKX phase was reported to be very sensitive to Mn deficiency \cite{STOu-Yang-2015,STOu-Yang-2014}, but the phase diagram observed for our sample exhibits no change compared to the literature for stoichiometric MnSi. This has also been confirmed via microscopic observation of the SKX via Small Angle Neutron Scattering experiments on a piece of the same sample\cite{SFobes-MnSiStrain}.

\begin{figure*}[htbp]
\hspace*{-10pt}
\vspace*{-30pt}
\includegraphics[width=17cm]{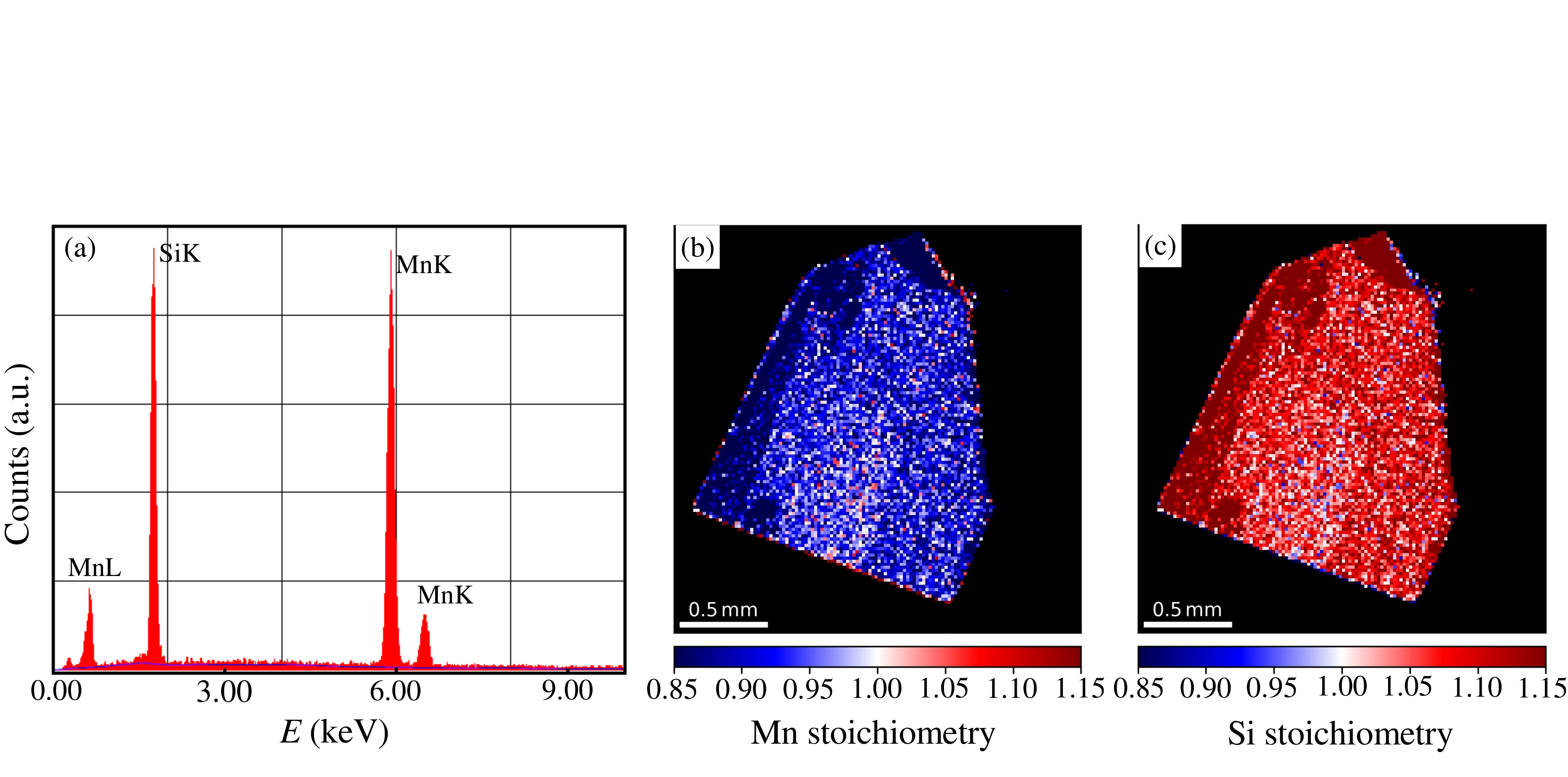}
\vspace*{-0pt}
\end{figure*}
\textbf{Fig.~S2 $|$ EDS characterization of MnSi.} (a) A representative EDS spectrum of MnSi. (b) and (c) show the maps of stoichiometric concentration of Mn and Si, respectively. See text for details.\\

\newpage

\textbf{SI \Rmnum{2}: $C\lowercase{_{ij}}$ \lowercase{and problem symmetry}}\\

For a 3D crystal, we write Hooke's law as\cite{SKittel-SolidState,SMigliori-RUS}
\begin{equation}
\boldsymbol{\sigma}=\mathbf{C}\cdot\boldsymbol{\varepsilon},
\label{EqS1}
\end{equation}
where $\boldsymbol{\sigma}$ is the stress tensor (6$\times$1), $\boldsymbol{\varepsilon}$ is the strain tensor (6$\times$1), and $\mathbf{C}$ is the elastic moduli matrix (6$\times$6). Expanding Eq.~(S1),
\begin{equation}
\left[
 \begin{array}{c}
    \sigma_{1} \\
    \sigma_{2} \\
    \sigma_{3} \\
    \sigma_{4} \\
    \sigma_{5} \\
    \sigma_{6} \\
  \end{array}
\right]=
\left[
  \begin{array}{cccccc}
    C_{11} & C_{12} & C_{13} & C_{14} & C_{15} & C_{16}  \\
    C_{21} & C_{22} & C_{23} & C_{24} & C_{25} & C_{26}  \\
    C_{31} & C_{32} & C_{33} & C_{34} & C_{35} & C_{36}  \\
    C_{41} & C_{42} & C_{43} & C_{44} & C_{45} & C_{46}  \\
    C_{51} & C_{52} & C_{53} & C_{54} & C_{55} & C_{56}  \\
    C_{61} & C_{62} & C_{63} & C_{64} & C_{65} & C_{66}  \\
  \end{array}
\right]\left[
  \begin{array}{c}
    \varepsilon_{1} \\
    \varepsilon_{2} \\
    \varepsilon_{3} \\
    \varepsilon_{4} \\
    \varepsilon_{5} \\
    \varepsilon_{6} \\
  \end{array}
\right].
\label{EqS2}
\end{equation}
Note that in this notation the subscripts 1 to 6 are actually two-element subscripts, and they correspond to the Cartesian coordinates by the following decoder table:
\begin{equation}
1\leftrightarrow xx; ~~~2\leftrightarrow yy; ~~~3\leftrightarrow zz; ~~~4\leftrightarrow yz,zy; ~~~5\leftrightarrow zx,xz; ~~~6\leftrightarrow xy,yx.
\label{EqS3}
\end{equation}
Because $C_{ij}$=$C_{ji}$, there are 21 different elastic moduli for an arbitrary system. The higher symmetry of the system, the fewer independent elastic moduli in $\mathbf{C}$. For cubic symmetry, $\mathbf{C}$ becomes
\begin{equation}
\mathbf{C}=\left[
 \begin{array}{cccccc}
    C_{11} & C_{12} & C_{12} & 0      & 0      & 0       \\
    C_{12} & C_{11} & C_{12} & 0      & 0      & 0       \\
    C_{12} & C_{12} & C_{11} & 0      & 0      & 0       \\
    0      & 0      & 0      & C_{44} & 0      & 0       \\
    0      & 0      & 0      & 0      & C_{44} & 0       \\
    0      & 0      & 0      & 0      & 0      & C_{44}  \\
  \end{array}
\right],
\label{EqS4}
\end{equation}
i.e., there are only three independent elastic moduli, $C_{11}$, $C_{12}$ and $C_{44}$. This is the case for MnSi in its paramagnetic state when $H$=0. The ground state of MnSi becomes a helimagnetic order below $T_c$, which in principle breaks the cubic symmetry. Treating MnSi as tetragonal for $H=0$ and $T<T_c$ is beyond the scope of this work and will be part of future efforts.

If we apply an external magnetic field $\textbf{H}$$\parallel$[001], the principal $\textbf{z}$-axis is no longer equivalent to $\textbf{x}$- and $\textbf{y}$-axes, the system can be regarded as tetragonal symmetry, and therefore, we have three extra elastic moduli, $C_{33}$, $C_{23}$(=$C_{31}$) and $C_{66}$. This situation is applicable to MnSi in the polarized paramagnetic state when a magnetic field is applied.

The skyrmion lines typically form a hexagonal lattice. An additional constraint $C^*$$\equiv$$(C_{11}$$-$$C_{12})/2$=$C_{66}$ is imposed to a hexagonal symmetry, requiring now only five independent elastic moduli. This is not what we observe experimentally in Fig.~3, as $C^*$$\neq$$C_{66}$, indicating that the SKX is not setting the symmetry of the system that remains tetragonal. Furthermore, as the SKX is coupled to the crystalline lattice, this enables the observation of changes measured in the elastic moduli as shown in Fig.~3. These changes are small compared with elastic moduli from the crystalline lattice, allowing to treat the system as one with tetragonal symmetry. However, how these two lattices with different symmetries couple to each other and how to obtain a more accurate determination of the elastic moduli of SKX from the measured bulk moduli require further investigations.

The situation becomes more complicated when a DC current is applied along [100]. The current is expected to break the tetragonal symmetry into an orthorhombic one. In principle, there should be nine independent elastic moduli, $C_{11}$, $C_{22}$, $C_{33}$, $C_{12}$, $C_{23}$, $C_{31}$, $C_{44}$, $C_{55}$ and $C_{66}$. This level of difficulty is not granted as the experimental results show that the applied current does not change the resonance frequencies or their field or temperature dependence, indicating that the symmetry of the system is not broken by low applied currents. \\

\textbf{SI \Rmnum{3}: D\lowercase{ata collection and analysis}}\\

In our RUS measurement, we swept frequency from 1250 to 5300 kHz. When frequency is approaching a vibrating mode of the sample, a resonance is detected. We show a representative RUS spectrum of MnSi in Fig.~S3 taken at $T$=34 K in the absence of magnetic field. For each resonance peak, the real part (in-phase, red) and imaginary part (out-of-phase, blue) of the signal have a Lorentzian shape $V({\it\omega})$=$z_0$$+$$A e^{i\phi}$/$({\it\omega}$$-$${\it\omega}_0$$+$$i{\it\Gamma}$$/$$2)$, where ${\it\Gamma}$ is the resonance width and $z_0$ is the background in the vicinity of the resonance. Plotting the real versus imaginary components of the resonance should form a circle in the complex plane (inset to Fig.~S3). Both the width (${\it\Gamma}$) and resonance frequency ($\omega_0$) were obtained following an established approach\cite{SShekhter-Pseudo}.

\begin{figure*}[htbp]
\includegraphics[width=15cm]{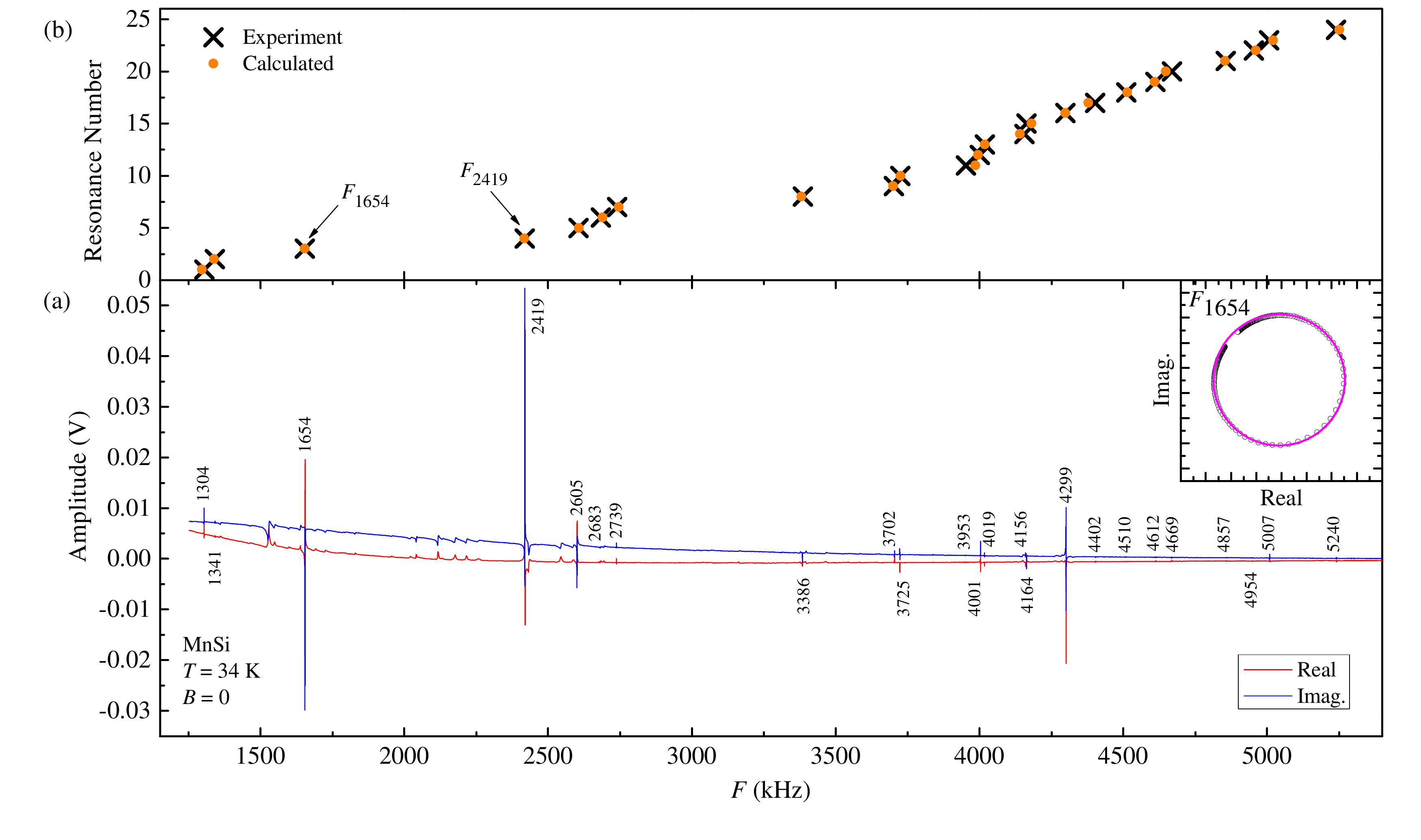}
\vspace*{-20pt}
\end{figure*}
\textbf{Fig.~S3 $|$ Full RUS spectrum of MnSi collected at $T$=34 K in the absence of a magnetic field.} The inset shows an Imaginary vs. Real plot of the resonance $F_{1654}$. The open symbols are experimental data, and the solid line defines a circle. \\

\begin{table*}
\caption{\label{TabS1} The least-square-fitting of RUS peaks at $T$=34 K and $H$=0. The sample dimensions are 1.446$\times$0.485$\times$0.767 mm$^3$. The fitting derives $C_{11}$=$C_{22}$=$C_{33}$=3.204192$\times$10$^2$ GPa, $C_{12}$=$C_{23}$=$C_{13}$=0.849543
$\times$10$^2$ GPa, and $C_{44}$=$C_{55}$=$C_{66}$=1.263798$\times$10$^2$ GPa. $\partial F/\partial C_{ij}$ (normalized) characterizes the dependencies of the fitting of a resonance peak on each $C_{ij}$.}
\begin{ruledtabular}
\begin{center}
\def\temptablewidth{1.6\columnwidth}
\begin{tabular}{ccccccccccccc}
$n$ & $F^{exp}$ (kHz) & $F^{cal}$ (kHz) & $Err$ (\%) &    &    &    &  \multicolumn{3}{l}{~~~~~~~~~$\partial F/\partial C_{ij}$}   &    &    &         \\
    &                 &                 &            & $C_{11}$ & $C_{22}$ & $C_{33}$ & $C_{23}$ & $C_{31}$ & $C_{12}$ & $C_{44}$ & $C_{55}$ & $C_{66}$ \\
\hline
  1 & 1304.090 & 1295.876 & -0.63 &  0.95 &  0.03 &  0.04 &  0.02 & -0.10 & -0.09 &  0.00 &  0.00 &  0.14 \\
  2 & 1341.150 & 1338.273 & -0.21 &  0.00 &  0.00 &  0.00 &  0.00 &  0.00 &  0.00 &  0.02 &  0.69 &  0.28 \\
  3 & 1654.028 & 1653.957 &  0.00 &  0.93 &  0.01 &  0.03 &  0.02 & -0.08 & -0.10 &  0.00 &  0.19 &  0.00 \\
  4 & 2418.830 & 2417.748 & -0.05 &  1.11 &  0.05 &  0.07 &  0.03 & -0.14 & -0.11 &  0.00 &  0.00 &  0.00 \\
  5 & 2605.364 & 2606.384 &  0.04 &  0.10 &  0.01 &  0.01 &  0.00 & -0.01 & -0.01 &  0.10 &  0.56 &  0.24 \\
  6 & 2682.732 & 2688.634 &  0.22 &  0.69 &  0.02 &  0.05 &  0.01 & -0.09 & -0.05 &  0.00 &  0.00 &  0.37 \\
  7 & 2739.155 & 2745.038 &  0.21 &  0.31 &  0.02 &  0.01 &  0.00 & -0.02 & -0.04 &  0.00 &  0.71 &  0.00 \\
  8 & 3385.540 & 3380.757 & -0.14 &  0.03 &  0.04 &  0.84 & -0.08 & -0.05 &  0.01 &  0.20 &  0.00 &  0.02 \\
  9 & 3702.542 & 3699.256 & -0.09 &  0.02 &  0.03 &  0.70 & -0.07 & -0.03 &  0.00 &  0.16 &  0.06 &  0.12 \\
 10 & 3724.552 & 3725.367 &  0.02 &  0.26 &  0.01 &  0.03 &  0.00 & -0.04 & -0.02 &  0.23 &  0.30 &  0.24 \\
 11 & 3953.041 & 3986.015 &  0.83 &  0.30 &  0.02 &  0.04 &  0.00 & -0.02 & -0.04 &  0.00 &  0.70 &  0.00 \\
 12 & 4001.499 & 3995.519 & -0.15 &  0.20 &  0.07 &  0.84 & -0.11 & -0.10 &  0.00 &  0.00 &  0.09 &  0.00 \\
 13 & 4018.697 & 4018.478 & -0.01 &  0.34 &  0.02 &  0.04 &  0.01 & -0.03 & -0.04 &  0.00 &  0.68 &  0.00 \\
 14 & 4155.843 & 4140.983 & -0.36 &  0.18 &  0.08 &  0.78 & -0.11 & -0.08 &  0.00 &  0.00 &  0.15 &  0.00 \\
 15 & 4163.504 & 4181.133 &  0.42 &  0.49 &  0.02 &  0.08 &  0.00 & -0.05 & -0.04 &  0.02 &  0.02 &  0.46 \\
 16 & 4299.043 & 4298.411 & -0.01 &  0.58 &  0.00 &  0.74 & -0.02 & -0.32 &  0.01 &  0.00 &  0.00 &  0.00 \\
 17 & 4402.400 & 4379.128 & -0.53 &  0.25 &  0.01 &  0.06 &  0.00 & -0.05 & -0.02 &  0.40 &  0.05 &  0.30 \\
 18 & 4510.335 & 4513.721 &  0.08 &  0.03 &  0.03 &  0.45 & -0.04 & -0.01 &  0.00 &  0.09 &  0.14 &  0.32 \\
 19 & 4612.002 & 4609.303 & -0.06 &  0.02 &  0.15 &  1.04 & -0.19 & -0.02 &  0.00 &  0.00 &  0.00 &  0.00 \\
 20 & 4669.281 & 4648.204 & -0.45 &  0.03 &  0.01 &  0.07 &  0.00 & -0.01 &  0.00 &  0.52 &  0.08 &  0.31 \\
 21 & 4856.554 & 4854.090 & -0.05 &  0.00 &  0.01 &  0.13 & -0.01 &  0.00 &  0.00 &  0.86 &  0.00 &  0.00 \\
 22 & 4953.862 & 4960.544 &  0.13 &  0.12 &  0.01 &  0.02 &  0.00 & -0.02 &  0.00 &  0.00 &  0.01 &  0.88 \\
 23 & 5007.325 & 5020.570 &  0.26 &  0.74 &  0.25 &  0.19 & -0.09 &  0.11 & -0.20 &  0.00 &  0.00 &  0.00 \\
 24 & 5240.128 & 5251.871 &  0.22 &  0.49 &  0.07 &  0.58 & -0.05 & -0.17 & -0.03 &  0.00 &  0.11 &  0.00 \\
\end{tabular}
\end{center}
\end{ruledtabular}
\end{table*}

For a sample with known elastic moduli, dimensions, and density, all the resonance peaks can be theoretically calculated by solving a 3D elastic wave function\cite{SMigliori-RUS,SMigliori-PhysicaB1993,SLeisure-RUS}. In our case, the elastic moduli are derived by a least-square-fitting, in which $C_{ij}$  are set as free fitting parameters. The iteration continues until $\chi^2$$\equiv$$\sum_{n}[(F_{n}^{cal}$$-$$F_n^{exp})/F_n^{exp}]^2$ minimizes, where $F_n^{exp}$ and $F_n^{cal}$ are the $n$th experimental and calculated frequencies, respectively. Table S1 shows an example of this fitting for a measurement at $T$=34 K and $H$=0 for cubic symmetry. The degree to which each resonance frequency depends on different $C_{ij}$, i.e. $\partial F/\partial C_{ij}$, is then calculated (normalized) and shown in the last nine columns of the table.\\

For this study, we have predominately tracked the two resonances $F_{2419}$ and $F_{1654}$ that are predominantly related to $C_{11}$. Although both resonances are suitable to track changes in the change of $C_{11}$ due to the presence of the skyrmion lattice, our measurements showed that the quality of $F_{1654}$ resonance peaks is better. In particular, the amplitude of $F_{2419}$ is very large (see Fig. S3), and therefore may reach the saturation limit of the pre-amplifier of the RUS setup. As a result the Imag. vs. Real plot no longer generates a good circle, which limits our analysis. This is avoided for the resonance $F_{1654}$ (cf the inset to Fig. S3). In turn, all measurements that were used to probe the depinning of the SKX as a function of applied current, were carried out using the resonance $F_{1654}$.\\

\textbf{SI \Rmnum{4}: T\lowercase{emperature dependence of $\uppercase{C}\lowercase{_{ij}}$}} \\

Figures S4(a-c) show the temperature dependencies of $C_{ij}$ of MnSi for $H$=0 that we used to define the phase boundaries. For the cubic symmetry of MnSi and in the absence of a magnetic field, only three independent elastic moduli are required. Here we follow the tradition of using $C_{11}$, $C_{12}$ and $C_{44}$ as the independent elastic moduli. In addition, the [110] shear modulus is defined as $C^{*}$$\equiv$$(C_{11}$$-$$C_{12})/2$.

\begin{figure*}[htbp]
\vspace*{-0pt}
\includegraphics[width=10cm]{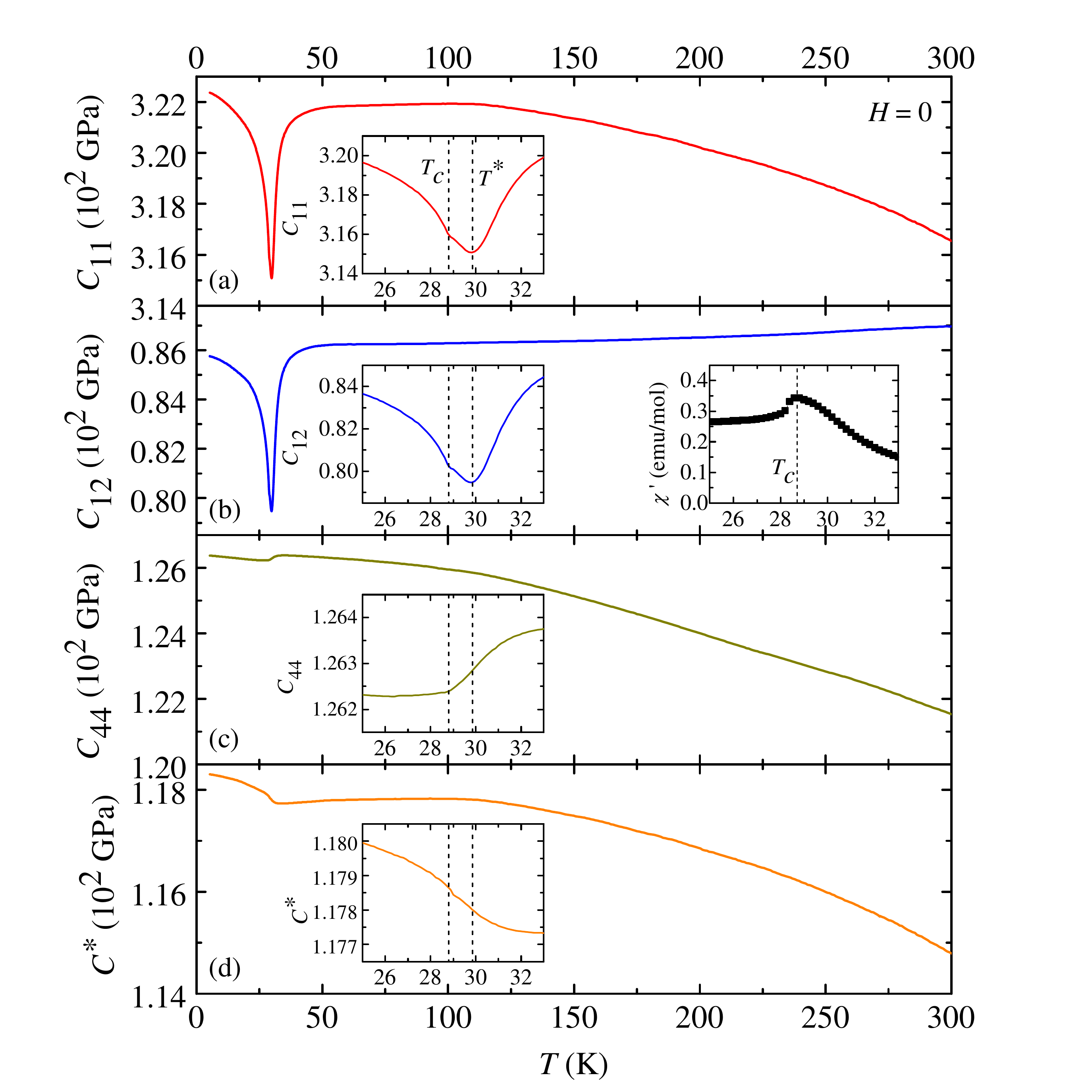}
\vspace*{-0pt}
\end{figure*}
\textbf{Fig.~S4 $|$  Temperature dependence of elastic moduli of MnSi.} (a-d) Zero-field temperature dependence of $C_{11}$, $C_{12}$ $C_{44}$ and $C^*$$\equiv$($C_{11}$$-$$C_{12}$)/2, respectively. The insets are near $T_c$=28.7 K. The right inset to panel (b) displays the real part of AC susceptibility $\chi'$ as a function of $T$. \\

At room temperature, we find $C_{11}$=316.54~GPa, $C_{12}$=86.96~GPa, and $C_{44}$=121.53~GPa in good agreement with the literature\cite{SPetrova-MnSiCij}. Upon cooling, both $C_{11}$ and $C_{44}$ increase. As temperature is lowered below 35 K, $C_{11}$ and $C_{44}$ drop but $C^{*}$ rises. This is characteristic of softening in one symmetry direction and stiffening in another in the vicinity of the magnetic phase transition to the helimagnetic (HM) state, highlighting the importance of measuring the full elastic tensor. The change of $C_{44}$ and $C^{*}$ are much smaller than those of $C_{11}$ and $C_{12}$. $C_{11}$ exhibits a minimum at $T^*$=29.8 K before recovering and then displaying an inflection at $T_c$=28.7 K where the system undergoes a PM-HM (helimagnetic) transition. The derived $T_c$ agrees well with AC susceptibility measurements shown in the right inset to Fig.~S4(b). In the low temperature limit, all $C_{ij}$ tend to saturate, as expected. It should be mentioned that $T^*$ is about 1 K above $T_c$, corresponding to a so-called Vollhardt invariance previously determined via specific heat and neutron scattering measurements\cite{SBauer-MnSi_FeCo,SJanoschek-MnSi2013}. Notably, $T^*$ is a characteristic temperature that describes a crossover from mean-field ferromagnetic to strongly-interacting HM fluctuations. Thus, the window between $T^*$ and $T_c$ describes the fluctuation-disordered (FD) region on the phase diagram [Fig.~2(b)], where strong critical fluctuations of the helical order parameter reduce both the correlation length and the mean-field helical phase transition temperature, resulting in a Brazovskii-type first-order transition at $T_c$ \cite{SJanoschek-MnSi2013,SBrazovskii-1975}. \\


%

\begin{thebibliography}{43}%

\bibitem{Courtney-MaterMechnical}
Courtney, T. H.  Mechanical behavior of materials. (Waveland Press, 2005).
\bibitem{Alava-2007}
Alava, M., Dub\'{e}, M. \& Rost, M. Imbibition in disordered media. {\it Adv. Phys.} \textbf{53}, 83-175 (2004).
\bibitem{Cayssol-DomainCreep}
Cayssol, F., Ravelosona, D., Chappert, C., Ferr\'{e}, J. \& Jamet, J. P. Domain Wall Creep in Magnetic Wires. {\it Phys. Rev. Lett.} \textbf{92}, 107202 (2004).
\bibitem{Larkin-JLTP1979}
Larkin, A. I. \&  Ovchinnikov, Y. N. Pinning in type II superconductors. {\it J. Low Temp. Phys.} \textbf{34}, 409-428 (1979).
\bibitem{blatterrev}
Blatter, G., Feigeloman, M. V., Geshkenbein, V. B., Larkin, A. I. \& Vinokur, V. M. Vortices in high-temperature superconductors. {\it Rev. Mod. Phys.} \textbf{66}, 1125-1388 (1994).
\bibitem{Lin-particle}
Lin, S. Z., Reichhardt, C., Batista, C. D. \& Saxena, A. Particle model for skyrmions in metallic chiral magnets: Dynamics, pinning, and creep. {\it Phys. Rev. B} \textbf{87}, 214419 (2013).
\bibitem{Muhlbauer-SKX}
M\"{u}hlbauer, S. {\it et al.} Skyrmion Lattice in a Chiral Magnet. {\it Science} \textbf{323}, 915-919 (2009).
\bibitem{Fert}
Fert, A., Cros, V. \& Sampaio, J. Skyrmions on the track. {\it Nat. Nanotech.} \textbf{8}, 152-156 (2013).
\bibitem{Mueller-2track}
M\"{u}ller, J. {\it et al.} Magnetic skyrmions on a two-lane racetrack. {\it New J. Phys.} \textbf{19}, 025002 (2017).
\bibitem{JiangW-HallCreep}
Jiang, W., {\it et al.} Direct observation of the skyrmion Hall effect. {\it Nat. Phys.} \textbf{13}, 162-169 (2017).
\bibitem{Reichhardt-SKXCreep}
Reichhardt, C. \& Reichhardt, C. J. O. Thermal creep and the skyrmion Hall angle in driven skyrmion crystals. {\it J. Phys.: Condens. Matter}
\textbf{31}, 07LT01 (2019).
\bibitem{Jonietz-MnSiCurrent}
Jonietz, F. {\it et al.} Spin Transfer Torques in MnSi at Ultralow Current Densities. {\it Science} \textbf{330}, 1648-1651 (2010).
\bibitem{Schulz-MnSiTHE}
Schulz, T. {\it et al.} Emergent electrodynamics of skyrmions in a chiral magnet. {\it Nat. Phys.} \textbf{8}, 301-304 (2012).
\bibitem{JZazvorka-2019}
Z\'{a}zvorka, J. \textit{et al.}, Thermal skyrmion diffusion used in a reshuffler device. \textit{Nat. Nanotech.} \textbf{14}, 658-661 (2019).
\bibitem{Yu-FeGe}
Yu, X. Z. {\it et al.} Skyrmion flow near room temperature in an ultralow current density. {\it Nat. Commun.} \textbf{3}, 988 (2012).
\bibitem{PinningSKX}
Reichhardt, C., Ray, D. \& Reichhardt, C. J. O. Collective Transport Properties of Driven Skyrmions with Random Disorder. {\it Phys. Rev. Lett.} \textbf{114}, 217202 (2015).
\bibitem{dong2015}
Dong, L. {\it et al.} Current-driven dynamics of skyrmions stabilized in MnSi nanowires revealed by topological Hall effect. {\it Nat. Commun.} \textbf{6}, 8217 (2015).
\bibitem{LuoY-MnSi2018}
Luo, Y. {\it et al.} Anisotropic magnetocrystalline coupling of the skyrmion lattice in MnSi. {\it Phys. Rev. B} \textbf{97}, 104423 (2018).
\bibitem{Anderson-RMP1964}
Anderson, P. W. \& Kim, Y. B. Hard Superconductivity: Theory of the Motion of Abrikosov Flux Lines. {\it Rev. Mod. Phys.} \textbf{36}, 39-43 (1964).
\bibitem{Janoschek-MnSi2013}
Janoschek, M. {\it et al.} Fluctuation induced first-order phase transition in Dzyaloshinskii-Moriya helimagnets. {\it Phys. Rev. B} \textbf{87}, 134407 (2013).
\bibitem{Nii-MnSiElastic}
Nii, Y., Kikkawa, A., Taguchi, Y., Tokura, Y. \& Iwasa, Y. Elastic Stiffness of a Skyrmion Crystal. {\it Phys. Rev. Lett.} \textbf{113}, 267203 (2014).
\bibitem{Petrova-MnSiSKX}
Petrova, A. E. \& Stishov, S. M. Field evolution of the magnetic phase transition in the helical magnet MnSi inferred from ultrasound studies. {\it Phys. Rev. B} \textbf{91}, 214402 (2015).
\bibitem{Migliori-RUS}
Migliori, A., \& Sarrao, J. L. Resonant Ultrasound Spectroscopy: Applications to Physics, Materials Measurements, and Nondestructive Evaluation. (Wiley, New York, 1997).
\bibitem{Migliori-PhysicaB1993}
Migliori, A. {\it et al.} Resonant ultrasound spectroscopic techniques for measurement of the elastic moduli of solids. {\it Physica B} \textbf{183}, 1-24 (1993).
\bibitem{Evans2017}
Evans, D. {\it et al.} Defect dynamics and strain coupling to magnetization in the cubic helimagnet Cu$_2$OSeO$_3$. {\it Phys. Rev. B} \textbf{95}, 094426 (2017).
\bibitem{Leroux-FeGeTHE}
Leroux, M. {\it et al.} Skyrmion Lattice Topological Hall Effect near Room Temperature. {\it Sci. Rep.} \textbf{8}, 15510 (2018).
\bibitem{Valenzuela2001}
Valenzuela, S. O. \& Bekeris, V. Oscillatory Dynamics and Organization of the Vortex Solid in YBa$_2$Cu$_3$O$_7$ Single Crystals. {\it Phys. Rev. Lett.} \textbf{86}, 504-507 (2001).
\bibitem{borisnatmat}
Maiorov, B. {\it et al.} Synergetic combination of different types of defect to optimize pinning landscape using BaZrO$_3$-doped YBa$_2$Cu$_3$O$_7$. {\it Nat. Mat.} \textbf{8}, 398-404 (2009).
\bibitem{campbellevettsnew}
Campbell, A. M. \& Evetts, J. E. Critical currents in superconductors. {\it Adv. Phys.} \textbf{50}, 1249-1449 (2001).
\bibitem{Valenzuela2002}
Valenzuela, S. O., Maiorov, B., Osquiguil, E. \& Bekeris, V. Elastic-to-plastic crossover below the peak effect in the vortex solid of YBa$_2$Cu$_3$O$_7$ single crystals. {\it Phys. Rev. B} \textbf{65}, 060504 (2002).
\bibitem{Bauer_MnSi_sampleshape}
Bauer, A. \& Pfleiderer C. Magnetic phase diagram of MnSi inferred from magnetization and ac susceptibility. {\it Phys. Rev. B} \textbf{85}, 214418 (2012).
\bibitem{Reimann-MnSinDI}
Reimann, T. {\it et al.} Neutron diffractive imaging of the skyrmion lattice nucleation in MnSi. {\it Phys. Rev. B} \textbf{97}, 020406(R) (2018).
\bibitem{Fobes-MnSiStrain}
Fobes, D. {\it et al.} Versatile strain-tuning of modulated long-period magnetic structures. {\it Appl. Phys. Lett.} \textbf{110}, 192409 (2017).
\end{thebibliography}

\begin{thebibliography}{10}%
\bibitem{SMuhlbauer-SKX}
M\"{u}hlbauer, S. {\it et al.} Skyrmion Lattice in a Chiral Magnet. {\it Science} \textbf{323}, 915-919 (2009).
\bibitem{SJanoschek-MnSi2013}
Janoschek, M. {\it et al.} Fluctuation induced first-order phase transition in Dzyaloshinskii-Moriya helimagnets. {\it Phys. Rev. B} \textbf{87}, 134407 (2013).
\bibitem{SBauer-MnSiC}
Bauer, A., Garst, M. \& Pfleiderer, C. Specific Heat of the Skyrmion Lattice Phase and Field-Induced Tricritical Point in MnSi. {\it Phys. Rev. Lett.} \textbf{110}, 177207 (2013).
\bibitem{SStishov-MnSi2007}
Stishov, S. M. {\it et al.} Magnetic phase transition in the itinerant helimagnet MnSi: Thermodynamic and transport properties. {\it Phys. Rev. B} \textbf{76}, 052405 (2007).
\bibitem{SBauer-MnSiPhase}
Bauer, A. \& Pfleiderer, C. Magnetic phase diagram of MnSi inferred from magnetization and ac susceptibility. {\it Phys. Rev. B} \textbf{85}, 214418 (2012).
\bibitem{SNii-MnSiElastic}
Nii, Y., Kikkawa, A., Taguchi, Y., Tokura, Y. \& Iwasa, Y. Elastic Stiffness of a Skyrmion Crystal. {\it Phys. Rev. Lett.} \textbf{113}, 267203 (2014).
\bibitem{SNewbury-EDS2013}
Newbury, D. \& Ritchie, N. W. M. Is Scanning Electron Microscopy/Energy Dispersive X-raySpectrometry (SEM/EDS) Quantitative? {\it SCANNING} \textbf{35}, 141-168 (2013).
\bibitem{SNewbury-EDS2015}
Newbury, D. \& Ritchie, N. W. M. Performing elemental microanalysis with high accuracy and highprecision by scanning electron microscopy/silicon drift detectorenergy-dispersive X-ray spectrometry (SEM/SDD-EDS). {\it J Mater. Sci.} \textbf{50}, 493-518 (2015).
\bibitem{STOu-Yang-2015}
Ou-Yang T. Y., Shu G. J., Lin J.-Y., Hu C. D. \& Chou F. C., Mn vacancy defects, grain boundaries, and A-phase stability of helimagnet MnSi. {\it J. Phys.: Condens. Matter} \textbf{28}, 026004 (2016).
\bibitem{STOu-Yang-2014}
Ou-Yang T. Y., Shu G. J., Hu C. D. \& Chou F. C., Manganese Deficiency in MnSi Single Crystal and Skyrmion Pinning. {\it IEEE Transactions on Magnetics} \textbf{50}, 150404 (2014).
\bibitem{SFobes-MnSiStrain}
Fobes, D. {\it et al.} Versatile strain-tuning of modulated long-period magnetic structures. {\it Appl. Phys. Lett.} \textbf{110}, 192409 (2017).
\bibitem{SKittel-SolidState}
Kittel, C. Introduction to Solid State Physics. (Wiley, New York, 2005).
\bibitem{SMigliori-RUS}
Migliori, A., \& Sarrao, J. L. Resonant Ultrasound Spectroscopy: Applications to Physics, Materials Measurements, and Nondestructive Evaluation. (Wiley, New York, 1997).
\bibitem{SShekhter-Pseudo}
Shekhter, A. {\it et al.} Bounding the pseudogap with a line of phase transitions in YBa$_2$Cu$_3$O$_{6+\delta}$. {\it Nature} \textbf{498}, 75-77 (2013).
\bibitem{SMigliori-PhysicaB1993}
Migliori, A. {\it et al.} Resonant ultrasound spectroscopic techniques for measurement of the elastic moduli of solids. {\it Physica B} \textbf{183}, 1-24 (1993).
\bibitem{SLeisure-RUS}
Leisure, R. G. \& Willis, F. A. Resonant ultrasound spectroscopy. {\it J. Phys.: Condens. Matter} \textbf{9}, 6001-6029 (1997).
\bibitem{SPetrova-MnSiCij}
Petrova, A. E. \& Stishov, S. M. Ultrasonic studies of the magnetic phase transition in MnSi. {\it J. Phys.: Condens. Matter} \textbf{21}, 196001 (2009).
\bibitem{SBauer-MnSi_FeCo}
Bauer, A. {\it et al.} Quantum phase transitions in single crystal Mn$_{1-x}$Fe$_x$Si and Mn$_{1-x}$Co$_x$Si: Crystal growth, magnetization, ac susceptibility, and specific heat. {\it Phys. Rev. B} \textbf{82}, 064404 (2010).
\bibitem{SBrazovskii-1975}
Brazovskii, S. A. Phase transition of an isotropic system to a nonuniform state. {\it Sov. Phys. JETP} \textbf{41}, 85-89 (1975).
\end{thebibliography}
\end{document}